# A strong-topological-metal material with multiple Dirac cones

Huiwen Ji[1], I Pletikosić[2,3], Q. D. Gibson[1], Girija Sahasrabudhe[1], T. Valla[3], R. J. Cava[1]

[1]*Department of Chemistry, Princeton University, Princeton, New Jersey 08544, USA*

[2]*Department of Physics, Princeton University, Princeton, New Jersey 08544, USA*

[3]*Condensed Matter Physics and Materials Science Department, Brookhaven National Laboratory, Upton, New York 11973, USA*

We report a new, cleavable, strong-topological-metal, $Zr_2Te_2P$, which has the same tetradymite-type crystal structure as the topological insulator $Bi_2Te_2Se$. Instead of being a semiconductor, however, $Zr_2Te_2P$ is metallic with a pseudogap between 0.2 and 0.7 eV above the fermi energy ($E_F$). Inside this pseudogap, two Dirac dispersions are predicted: one is a surface-originated Dirac cone protected by time-reversal symmetry (TRS), while the other is a bulk-originated and slightly gapped Dirac cone with a largely linear dispersion over a 2 eV energy range. A third surface TRS-protected Dirac cone is predicted, and observed using ARPES, making $Zr_2Te_2P$ the first system to realize TRS-protected Dirac cones at $\bar{M}$ points. The high anisotropy of this Dirac cone is similar to the one in the hypothetical Dirac semimetal $BiO_2$. We propose that if $E_F$ can be tuned into the pseudogap where the Dirac dispersions exist, it may be possible to observe ultrahigh carrier mobility and large magnetoresistance in this material.

## I. INTRODUCTION

3D Dirac materials that host massless Dirac fermions have recently received widespread attention due to their exotic properties such as extremely high carrier mobility and large magnetoresistance.[1-3] They are the 3D analogues of graphene, with Dirac points and "photon-like" dispersions in their electronic structures. The Dirac bands can have different origins. Breaking their protecting symmetries leads the Dirac point to be gapped and a Dirac band where the electrons have mass. Topological insulators such as $Bi_2Te_3$ and $Bi_2Se_3$ have a finite bulk band gap due to spin-orbit coupling (SOC) and chiral, massless Dirac surface states.[4,5] In these materials, the degeneracy at the surface Dirac point is protected by time-reversal symmetry (TRS).[6,7] The Dirac points are doubly degenerate and the corresponding surface states are strongly spin-polarized. By contrast, another group of 3D Dirac materials – Dirac semimetals – have Dirac dispersions in the bulk electronic structure with Dirac points that are protected by crystalline symmetries.[8] Their Dirac points are 4-fold degenerate and do not have spin polarization. Recently established Dirac semimetals include $Cd_3As_2$[2,9,10] and $Na_3Bi$.[3,11]

Given the potential applications and intrinsically novel physics of 3D Dirac materials, new candidates are wanted to diversify the current family. Among different material systems, layered chalcogenides containing heavy elements are favored for having the band degeneracies necessary for either 3D Dirac dispersions or topologically protected surface states. Their layered structures and easy cleavage allow visualization of any Dirac dispersion that may be present through ARPES (Angle-resolved Photoemission Spectroscopy) study. The Tetradymite structure type is of particular importance as it hosts many 3D topological insulators, namely $Bi_2Te_3$, $Bi_2Se_3$, $Sb_2Te_3$, $Bi_2Te_2Se$ and $Bi_2Te_{1.6}S_{1.4}$.[5,12,13] These materials all have a single 2D Surface Dirac cone in the Brillouin zone. Here, we report a new strong-topological-metal, $Zr_2Te_2P$, in the Tetradymite structure family. Its multiple Dirac features, which originate from both the bulk and the surface, are theoretically predicted and experimentally observed using ARPES. In contrast with the majority of currently known topological materials, which are based on *s-p* band inversions, the band inversions in $Zr_2Te_2P$ are of the *d-p* type.

## II. EXPERIMENT

In this work, single crystals of both pristine $Zr_2Te_2P$ and slightly Cu-intercalated $Cu_{0.06}Zr_2Te_2P$ were grown through vapor transport.[14] Initially, adding copper was intended to increase the $E_F$ of pristine $Zr_2Te_2P$ because copper intercalation in layered chalcogenides has been known to inject electrons while maintaining the overall band structure.[15,16] We could only intercalate 0.06 Cu per formula unit, (determined using single crystal X-ray diffraction, see TABLE I in the Supplemental Material[17]), i.e. 0.03 electrons per Zr, which can raise $E_F$ only minimally in this metallic material. We found, however, that the as-grown Cu intercalated crystals have higher quality, and therefore the experimental characterizations were performed on $Cu_{0.06}Zr_2Te_2P$, which we designate as $Cu-Zr_2Te_2P$ in the following.

The sample synthesis followed the methods reported elsewhere.[14] Polycrystalline samples of $Zr_2Te_2P$ and $Cu-Zr_2Te_2P$ were made through traditional solid state synthesis at 1000℃. The starting formulas were $Zr_2Te_2P$ and $Cu_{0.8}Zr_2Te_2P$, respectively. Careful control of the temperature and close monitoring was required to avoid explosion due to phosphorus' high vapor pressure and the large enthalpy of formation of the product. The polycrystalline powder was then loaded into a quartz ampoule containing iodine as a transport agent. A temperature gradient for chemical vapor transport was created by using a horizontal two-zone furnace with the source end set at 800℃ and the sink end at 900℃ for five days. Scattered silvery hexagonal platelets formed at the hot end. The layered crystal structure of $Cu_{0.06}Zr_2Te_2P$ was confirmed by using PXRD (Powder X-ray Diffraction). Single crystals of $Cu_{0.06}Zr_2Te_2P$, were selected and studied by using a Photon 100 single crystal X-ray diffractometer with a Mo radiation source.

The ARPES maps over a two-dimensional momentum space were acquired at the HERS endstation of the Advanced Light Source (Berkeley, CA) by rotating the Scienta R4000 electron analyzer around the axis parallel to the direction of its entrance aperture. Photons of 50 eV were used as the excitation. The resolution was set to 25 meV in energy, and 0.2° and 0.5° in the angles. The samples were cleaved and kept at 15 K in ultrahigh vacuum in the course of the measurements.

The electronic structure calculations were performed in the framework of density functional theory (DFT) using the WIEN2K code with a full-potential linearized augmented plane-wave and local orbitals [FPLAPW + lo] basis[18-20] together with the Perdew Burke Ernzerhof (PBE) parameterization of the generalized gradient approximation (GGA) as the exchange-correlation functional. The plane wave cut-off parameter $R_{MT}K_{MAX}$ was set to 7, and the reducible Brillouin zone was sampled by 2000 k-points. The lattice parameters and atomic positions employed for the structural input are from Ref. 14. For the slab calculation for surface electronic structure simulation, a $10 \times 10 \times 1$ k-mesh was used.

## III. RESULTS AND DISCUSSION

As shown in Fig. 1(a), $Zr_2Te_2P$ crystalizes in the same structure as the well-known topological insulator $Bi_2Te_2Se$,[21] with a rhombohedral space group $R\bar{3}m$ (No. 166). Each unit cell is composed of three quintuple atomic layers, which are separated from each other by van der Waals gaps. In both crystal structures, the more electropositive element (Zr or Bi) takes the 6c Wyckoff position and adopts octahedral coordination. Like $Bi_2Te_2Se$, $Zr_2Te_2P$ has a fully ordered crystal structure with distinct Wyckoff positions for P and Te (see TABLE I in Supplemental Material). A typical SEM image of a single crystal piece is shown in Fig. 1(c). The crystal surface was exfoliated with Scotch Tape before imaging, exposing a microscopically smooth face under SEM. Its 120-degree edges are a reflection of the hexagonal close packing of atoms within the layers. The Miller index (001) of the basal plane crystal surface is confirmed by an X-ray diffraction pattern taken on a piece of single crystal (Fig. 1(b)), where only (003$n$) reflections are visible.

The calculated bulk electronic structure of $Zr_2Te_2P$, with spin orbit coupling included, is shown in Fig. 2; the metallic features agree well with previous transport measurements on this compound.[14] The three most important bands near $E_F$ are denoted by black arrows and the labels Band 1, 2 and 3. The band dispersions are plotted along four time-reversal invariant momenta (TRIM). Several flatly dispersive bands observed along the $\Gamma - Z$ direction originate from the poor orbital overlap along the $c$ axis of the structurally two-dimensional compound. Nevertheless, there remain some steep dispersions along the $\Gamma - Z$ direction due to the hybridization of Zr $4d_{z^2}$ orbitals and Te $5p_z$ orbitals. We observe a pseudogap between the energies of 0.2 and 0.7 eV above $E_F$ at the $\Gamma$ point (between Band 1 and Band 2), which agrees closely with a strong suppression of the density of states (DOS) in the calculation (Fig. S2 in Supplemental Material). In addition, a rather flat band is observed at around $-3.25$ eV below $E_F$, which mostly comes from P 3$p$-orbitals. As expected, sharp peaks in the DOS from P are observed in the DOS calculation (Fig. S2). Because of the large electronegativity difference between Zr and P, the bonded electrons are localized with low mobility.

When SOC is included in the DFT calculations, a band inversion is predicted at the Γ point (highlighted by a red circle in Fig. 2(a)). This band inversion happens between the Zr 4$d$ orbitals and the Te 5$p$ orbitals. The band dispersions in Fig. 2(b) and (c) are Zr 4$d$-orbital and Te 5$p$-orbital characteristic plots, respectively. As shown, the Zr 4$d$-orbital dominates the entire conduction band above the pseudogap while vanishing at the tip; the Te 5$p$-orbital acts oppositely. These observations imply a band inversion between the two orbitals. The surface states that originate due to this band inversion (a Dirac cone) are topologically nontrivial according to the parity analysis of the top-most isolated bands below Band 1 (bands buried deeper are ignored as they do not alter the overall topology) at the four TRIM. The parities at Γ, F, L, Z points are positive, positive, positive, and negative, respectively, leading to a nontrivial $Z_2$ topological quantum number of $v_0 = 1$. The nonzero $Z_2$ guarantees the existence of topological surface states on all crystal faces[6]. If $E_F$ can be tuned through chemical doping or gating into the pseudogap, then $Zr_2Te_2P$ would become a strong topological metal (not a topological insulator as there are a few band dispersions along low-symmetry momenta that will be present). This band inversion is unlike the ones in traditional topological insulators such as $Bi_2Te_3$ and $Bi_2Se_3$ where only $p$ orbitals are involved (between Bi 6$p$ and Te 5$p$ (or Se 4$p$)) because it involves metal $d$ states and Te $p$ states. The interaction between electrons is predicted to be stronger in $d$-$p$ topological insulators because the relatively localized nature of $d$ orbitals increases the effective mass of bulk electrons, which in turn favors the surface conductivity.[22]

Here, one might wonder about the bulk states co-existing in the pseudogap with the surface states, since they would affect the transport properties of $Zr_2Te_2P$ even if $E_F$ is tuned successfully into the pseudogap. Although invisible between the TRIM, when the rhombohedral Brillouin zone is folded into a hexagonal one, the calculations show a clear bulk-derived Dirac cone between M and K momenta (Fig. S1 (left)). It spans an energy range of over 2 eV. The bands at the Dirac point have a quadruple degeneracy. However, because the band crosses at low-symmetry points, neither TRS nor crystalline symmetry protects it from developing a band gap due to SOC. Thus, as expected, when SOC is considered, the degeneracy at the Dirac point is lifted and a tiny gap of around 70 meV is formed (Fig. S1(right)), although the remainder of the linear dispersion of the Dirac cone is intact. This kind of gapped Dirac point leads to Dirac states with mass (however, in this case, the gap, and therefore the mass, is very small). We thus speculate that if $E_F$ can be tuned into the pseudogap through chemical doping or gating in $Zr_2Te_2P$, a surface-originated spin-polarized Dirac cone and a bulk-originated spin-degenerate Dirac dispersion would come into play along with few other bulk states. Although the transport will be dominated by bulk conduction, the Dirac-like dispersion of these bands might lead to high carrier mobility comparable to other topological semimetals such as $Cd_3As_2$.

The bulk calculations that suggest the presence of a topologically-protected Dirac cone on the surfaces of $Zr_2Te_2P$ at the $\Gamma$ point agree closely with the slab calculation shown in Fig. 3(a), where an artificial unit cell is constructed by stacking five quintuple layers of Te-Zr-P-Zr-Te and putting a vacuum space at the two ends. The slab calculation therefore predicts the surface electronic structure on the (001) crystal face. Accordingly, the rhombohedral bulk Brillouin zone is projected onto a hexagon-shaped surface Brillouin zone (Fig. 3(b)) where the four TRIM collapse into $\bar{\Gamma}$ and $\bar{M}$. In the slab calculation, the band dispersions are plotted with circles whose size is proportional to the contribution from the surface atomic layer. A linear dispersion of surface states is observed inside the pseudogap with the band degeneracy/Surface Dirac point (SDP, highlighted by a red circle) occurring at around 0.4 eV above $E_F$. Unfortunately, these calculated surface states are difficult to visualize through ARPES as they sit above $E_F$, and only occupied states are visible by that technique. Integration of the DOS suggests that 0.6 – 0.8 electrons per formula unit are needed to dope the system into the pseudogap.

Nevertheless, in the slab calculations, a third Dirac-cone-like feature is found at the $\bar{M}$ point at around – 0.8 eV below $E_F$ (highlighted by a green circle in Fig. 3(a)). It has linear dispersions around it in a 0.4 eV energy window. This cone was absent in the bulk electronic structure calculation, and thus originates from the surface. Meanwhile, the fact that the cone locates at the $\bar{M}$ point, on the edge of the first surface Brillouin zone, indicates that there are three such cones in the first surface Brillouin zone. The two bands are degenerate at the time-reversal-invariant $\bar{M}$ while the degeneracy is lifted away from $\bar{M}$. This Dirac cone is highly anisotropic, similar to the Dirac cone predicted for the hypothetical topological Dirac semimetal $BiO_2$[8]. It has photon-like massless dispersions along $\bar{M} \to \bar{K}$ but is almost flat along $\bar{\Gamma} \to \bar{M}$ (meaning the electron velocity is almost zero). Notice that the two Dirac bands along $\bar{M} \to \bar{\Gamma}$ are so close to each other that they almost form a Dirac line.

Recently it has been shown that even strong topological insulators with a non-trivial parity invariant can have gapped surface states if away from TRIM points.[23] Although $\bar{M}$ is a time-reversal-invariant point, $\bar{K}$ is not. Therefore, to analyze the topological nature of this Dirac-cone, a parity analysis similar to that performed for the Dirac cone at Γ was performed regarding the gap between Band 2 and Band 3 (Fig. 2(a)). The $Z_2$ topological quantum number $\nu_0$ was again calculated to be 1 (positive parity at Γ, Z, and F while negative at L), making the surface states along $\bar{\Gamma} - \bar{M}$ robust and visible on any arbitrary crystal termination. However, given that $\bar{K}$ is not a TRIM, the surface state along $\bar{M} - \bar{K}$ cannot be protected solely by TRS. In this case the crystal symmetry adds another constraint. In the slab calculation, the full symmetry of the slab at the $\bar{K}$ point is $D_3$. This double group has three different irreducible representations. Two of them have a $C_3$ eigenvalue of − 1 and mirror eigenvalues of +/− i, which map to spin and are not constrained to be degenerate due to the non-TRIM nature of the $\bar{K}$ point. The other irreducible representation has a $C_3$ eigenvalue of + 1 and a mirror eigenvalue of 0, and is doubly degenerate, regardless of the non-TRIM nature of the $\bar{K}$ point. This implies that doubly degenerate states, from crystal symmetry, are possible even at the non-TRIM $\bar{K}$ point, thus allowing for topologically-protected states along the $\bar{M} - \bar{K}$ direction. Due to the $C_3$ symmetry at $\bar{K}$, some bands are doubly degenerate. This would have the result of pinning the states into doublets at the $\bar{K}$ point, thus making the surface state along $\bar{M} - \bar{K}$ protected by both TRS and $C_3$ symmetries. Breaking either of these symmetries would result in a trivial surface state along the $\bar{M} \to \bar{K}$ direction. It is worth noting, however, that breaking $C_3$ symmetry would preserve the surface states along $\bar{\Gamma} - \bar{M}$ as that is protected solely by TRS.

ARPES measurements were employed to experimentally probe the electronic structure of Cu-Zr$_2$Te$_2$P. The ARPES intensity mappings along different directions denoted in Fig. 3(c) (Fig. 3(c) is an ARPES mapping of Fermi surface) are shown in Fig. 3(d) – (f). The high quality of the Cu-Zr$_2$PTe$_2$ crystals employed is manifested in the sharp ARPES spectra. The third Dirac cone predicted by the DFT calculations is clearly observed (Fig. 3(e)) with a large energy window (over 2 eV) of linear dispersions around it. Its Dirac point is observed at the $\bar{M}$ point at around − 1.0 eV, which agrees closely with the calculation. The cuts along $\bar{\Gamma} \to \bar{M}$ and $\bar{\Gamma} \to \bar{K}$ directions (Fig. 3(d) and (f)) also match with the slab calculation. For instance, the cutoff energy ($E_F$) measured in the intensity mapping along the $\bar{\Gamma} \to \bar{K}$ direction is at the top of a highly dispersive band, confirming the calculation along the same direction. Also, along the $\bar{\Gamma} \to \bar{M}$ direction, the bands are more dispersive near $\bar{\Gamma}$ while becoming flat when close to $\bar{M}$. Both features are manifested in Fig. 4(d) taken along the same direction.

Despite the exotic Dirac features both inside the pseudogap above $E_F$ and at around $-1.0$ eV below $E_F$, the electronic properties of bulk $Zr_2Te_2P$ are dominated by the electronic dispersions near $E_F$. In Fig. 2(a), the dispersive band stretching from the close proximity of the $\Gamma$ point along the $\Gamma - F$ and $\Gamma - L$ directions suggests the existence of a dominant electron pocket at $E_F$. In addition, close to the $\Gamma$ point, several bands cross $E_F$ and form circular hole-like fermi surfaces. These two kinds of Fermi surfaces are clearly observed in the calculated bulk Fermi surface in Fig. 4(a), which is viewed down the $\Gamma - Z$ direction. Due to the three-fold rotational and inversion symmetries of the compound, there are six ellipsoidal electron pockets around $\Gamma$. These electron pockets in fact host the Dirac cones at the $\bar{M}$ points. The hole pockets centered at the $\Gamma$ point are also displayed in Fig. 4(a), in different colors from the electron pockets.

The experimental Fermi surface obtained from the ARPES characterization is shown in Fig. 4(b). It shows the electron and hole pockets that are predicted by the DFT calculation. The dotted lines in the image outline the surface Brillouin zone. The plum-pit-shaped electron pockets along $\bar{\Gamma} \to \bar{M}$ suggest that the surface-projected electronic structure is highly anisotropic. Fig. 4(c) shows a stack of equal- energy-spaced constant-energy ARPES contours. The topmost map is at $E_F$ while the bottom one is at $-1.1$ eV from $E_F$. The white lines are drawn as a guide to the eye. The electron pocket narrows as the binding energy increases. The pocket shrinks linearly along the transverse direction ($\bar{M} \to \bar{K}$) and vanishes at $\bar{M}$ point, which corresponds to the Dirac point. Along the longitudinal direction ($\bar{\Gamma} \to \bar{M}$), however, the pocket first shrinks linearly but more drastically when close to the Dirac point. This is consistent with the dispersion along $\bar{\Gamma} \to \bar{M}$ (Fig. 3(d)) which is quite linear immediately down the $E_F$ while flattens out in the close proximity of the Dirac point.

In-situ deposition in the ARPES chamber of a small amount of potassium on the surface of $Zr_2Te_2P$ was employed to electron-dope of the system so that normally unoccupied bands could be observed. With the bulk electronic structure almost intact, the potassium deposition we performed simply shifts the $E_F$ upward by around $100 - 200$ meV. Although the amount of doping was insufficient to observe the Dirac surface states predicted to exist inside the pseudogap, it revealed a new Dirac feature near $E_F$ (Fig. 4(e)). (The $E_F$ in Fig. 4(d) and (e) represents the Fermi energy of n-doped $Zr_2Te_2P$, not the $E_F$ of the pristine compound in Fig. 4(b) and (c)). From $-0.2$ eV to $E_F$, the electron pockets begin to touch each other at a point on the $\bar{\Gamma} - \bar{K}$ azimuth. Here, the red dotted line in Fig. 4(d) points to an azimuth that is parallel to the $\bar{M} \to \bar{K}$ direction. The energy-dependent mapping along this direction is shown in Fig. 4(e). Two Dirac-nodes are observed close to the new E$_F$. Since the two Dirac nodes are away from TRIM points and are not protected by any crystalline symmetry, spin-orbit coupling results in gapped Dirac points. We deduce that these massive Dirac nodes come from the slightly gapped bulk-derived Dirac dispersion inside the pseudogap (Fig. S1(right)). The Dirac nodes are posited at 0.6 eV above the pristine E$_F$ between M and K but lower in energy when travelling away from M – K towards Γ, leading to a Dirac node near the new E$_F$ (shown in Fig. 4(e)). Another Dirac-like crossing at around – 1.0 eV between the two massive Dirac nodes in Fig. 4(e) corresponds to the extension of the Dirac point along $\bar{\Gamma} \to \bar{M}$. This crossing is not gapless due to the tiny splitting of the dispersion along $\bar{\Gamma} \to \bar{M}$ from the Dirac point. These bands therefore also connect to the surface states centered around $\bar{M}$ as discussed earlier.

## IV. CONCLUSION

In conclusion, we have reported the discovery of a new strong-topological-metal, Zr$_2$Te$_2$P, where the band inversions are of the *d-p* type. Zr$_2$Te$_2$P adopts the same crystal structure as the topological insulator Bi$_2$Te$_2$Se. It has two TRS-protected surface Dirac cones and one gapped 3D Dirac cone in its electronic structure. The Dirac cone at the $\bar{\Gamma}$ point sits in a pseudogap at around 0.4 eV above E$_F$, which is a result of the band inversion between Zr 4*d* orbitals and Te 5*p* orbitals. A second Dirac cone at the $\bar{M}$ point is derived from an electron pocket at E$_F$, which shrinks when lowering in energy and vanishes to a doubly degenerate Dirac point at – 1.0 eV. This Dirac cone is unusually anisotropic, with a linear dispersion along $\bar{M} \to \bar{K}$ but an almost flat, nearly degenerate dispersion along $\bar{\Gamma} \to \bar{M}$, similar in some ways to the surface states observed in Ru$_2$Sn$_3$,[24] the nature of which is still not understood. Unfortunately, the physical properties due to the second Dirac cone are unlikely to be isolated due to the coexistence of bulk bands in the same energy range. Nevertheless, this is the first system to realize a TRS-protected Dirac cone at $\bar{M}$ point although that was predicted for the (111) surface of the Bi-Sb alloy system.[25] When shifting E$_F$ upward through electron-doping, the Dirac pockets touch each other on the $\bar{\Gamma} - \bar{K}$ azimuth, forming a massive Dirac node gapped by spin-orbit coupling. We propose that if E$_F$ can be shifted into the pseudogap, it is possible to realize high electron mobility and strong anisotropy in this system. Both chemical doping and gating are promising methods to realize this goal.

# ACKNOWLEDGEMENTS

This work was supported by the ARO MURI on topological insulators, grant W911NF-12-1-0461. The ARPES measurements at the Advanced Light Source were supported by the U.S. Department of Energy under Contract Nos DE-AC02-05CH11231 and DE-SC0012704. The authors thank Lukas Müchler, Liang-Yan Hsu and Bin Liu for helpful discussions.

## CAPTIONS

**FIG. 1** (Color online) (a) The crystal structure of $Zr_2Te_2P$. (b) X-ray diffraction pattern from the basal plane crystal surface of Cu-$Zr_2Te_2P$. (c) SEM image of a typical Cu-$Zr_2Te_2P$ single crystal.

**FIG. 2** (Color online) (a) The bulk band structure of $Zr_2Te_2P$. The band inversion is highlighted with a red circle, which is further characterized in the Zr 4$d$-orbital (b) and Te 5$p$-orbital (c) characteristic plots. The thickness of the bands indicates the contribution from a specific orbital.

**FIG. 3** (Color online) (a) The slab electronic structure calculation that simulates the surface electronic structure. The band dispersions are plotted with circles whose size is proportional to the contribution from the surface atomic layer. Two topologically-protected Dirac cones are highlighted with a red and a green circles. (b) The rhombohedral Brillouin zone projects onto a hexagonal surface Brillouin zone. All the TRIM are labeled. (c) Experimental ARPES intensity map of the Fermi surface. The green arrows denote the high-symmetry directions along which the dispersion of Cu-$Zr_2Te_2P$ bands was determined by ARPES. These dispersions are shown in Fig. (d), (e) and (f), which are cuts along $\bar{\Gamma} - \bar{M}$, $\bar{K} - \bar{M} - \bar{K}$, and $\bar{K} - \bar{\Gamma} - \bar{K}$ directions, respectively.

**FIG. 4** (Color online) (a) DFT-calculated Fermi surface viewed down the $\Gamma - Z$ direction in the first Brillouin zone. (b) Intensity mapping of the experimental Fermi surface obtained from ARPES. The dotted lines in the image outline the boundaries of surface Brillouin zone. (c) A stack of experimentally determined equal-energy-spaced constant-energy ARPES contours. The white lines are drawn as a guide to the eye. (d) The petal-like electron pockets were made to touch by electron doping of the bands. Constant energy cuts show the shift of the band structure by some 150 meV. The red dotted line denotes where the electron pockets begin to touch. An energy-dependent mapping along this direction (parallel to the $\bar{M} \to \bar{K}$) is shown in (e).

**FIG. S1** (Color online) Calculated band structures of $Zr_2Te_2P$ with (right) and without (left) spin-orbit coupling. A Dirac point is observed inside the pseudogap above $E_F$ without the consideration of spin-orbit coupling. However, a small gap of ~ 70 meV is opened when spin-orbit coupling is taken into account; even with the SOC the majority of these Dirac bands remain linear over a 2 eV energy range.

**FIG. S2** (Color online) Density of states calculation for $Zr_2Te_2P$. A pseudogap is observed at 0.2 – 0.7 eV above $E_F$.

**FIG. 1**

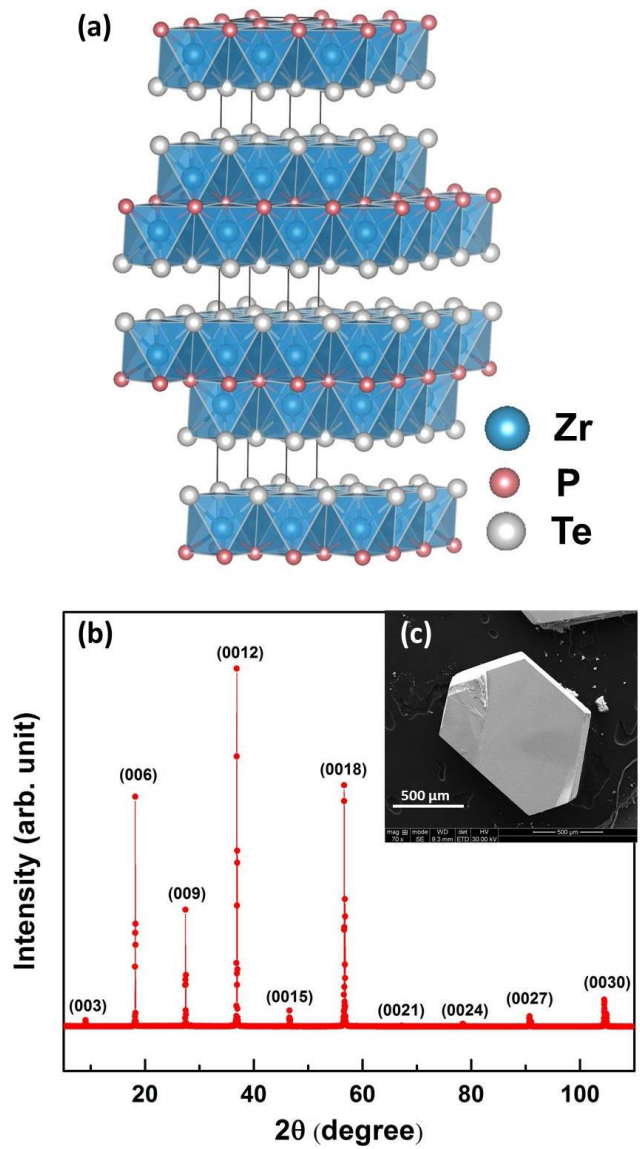

**FIG. 2**

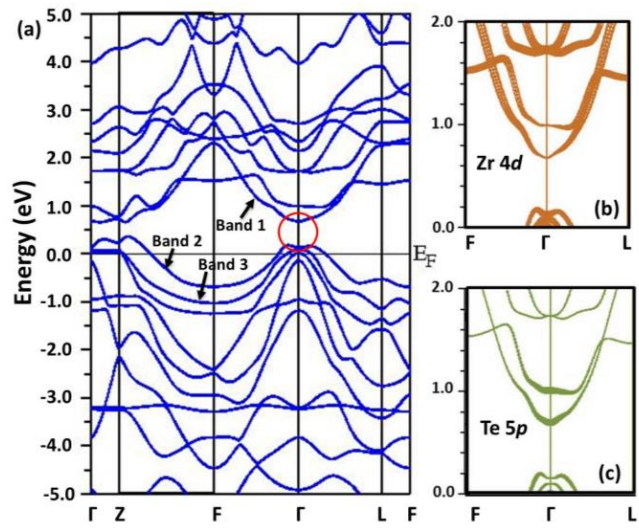

**FIG. 3**

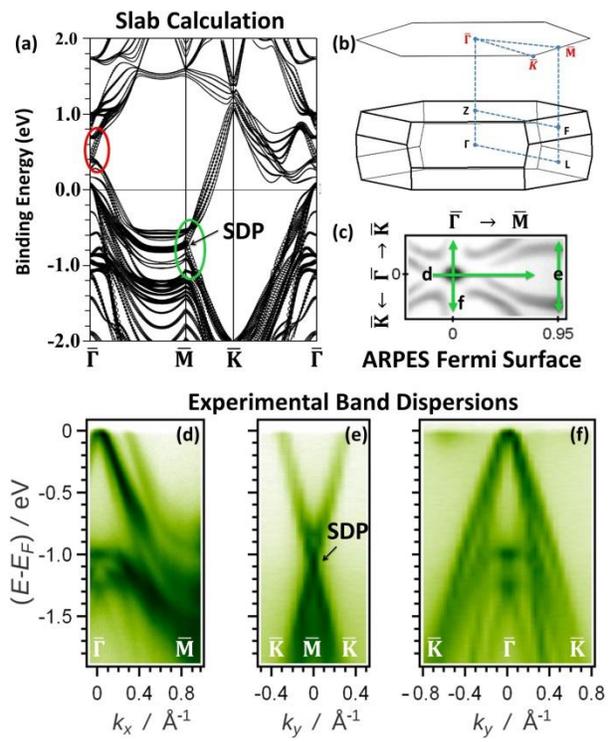

FIG. 3. (a) Slab calculation. (b) Brillouin zone. (c) ARPES Fermi Surface. (d–f) Experimental Band Dispersions.

**FIG. 4**

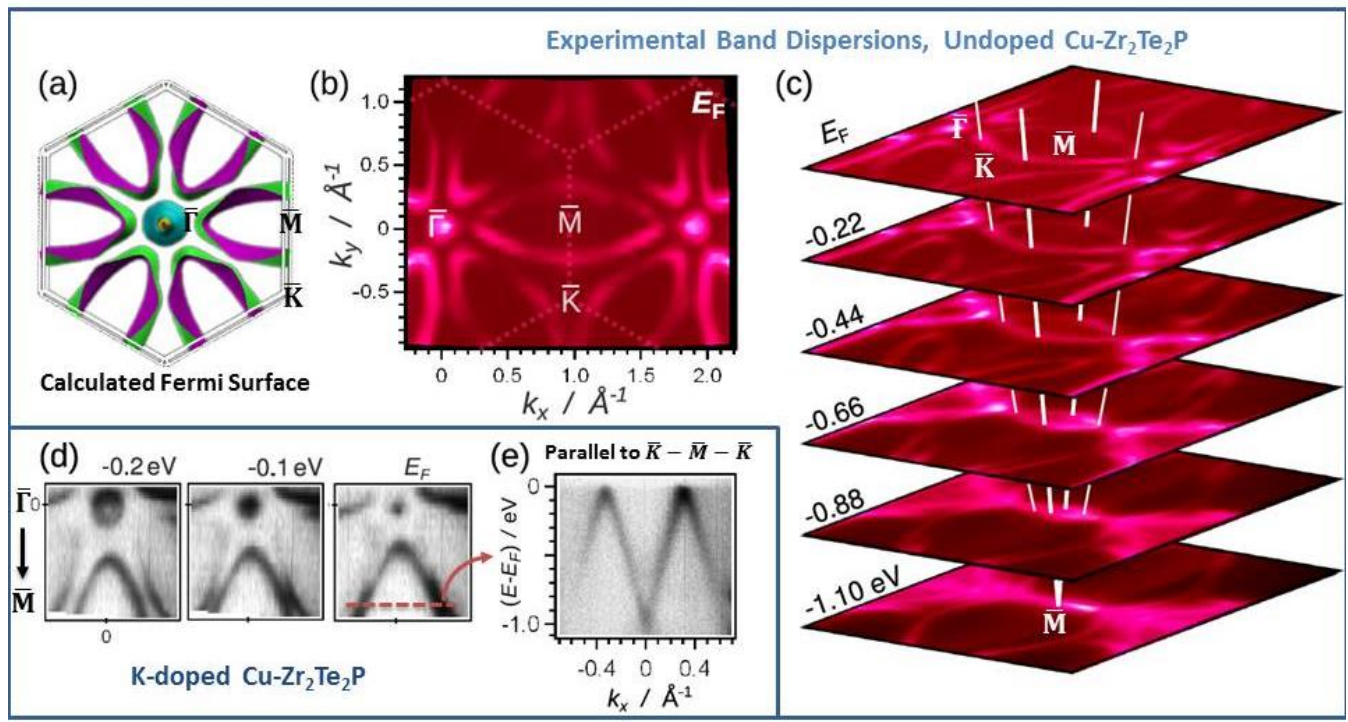

# SUPPLEMENTAL MATERIAL

TABLE I. Structural characterization of $Cu_{0.06}Zr_2Te_2P$. Space group $R\bar{3}m$ (No. 166), $a$ = 3.8202(2) Å, $c$ = 29.2085(19) Å, 240 unique reflections, $R_1$ (all reflections) = 0.0307, $wR_2$ = 0.0516, Goodness of fit = 1.179.

| Atom | Wyck. | $x$ | $y$ | $z$ | Occ. | $U_{11}$ | $U_{22}$ | $U_{33}$ |
|---|---|---|---|---|---|---|---|---|
| Zr | 6c | 0 | 0 | 0.38261(2) | 1 | 0.0069(2) | 0.0069(2) | 0.0090(3) |
| Te | 6c | 1/3 | 2/3 | 0.44684(14) | 1 | 0.00800(16) | 0.00800(16) | 0.0093(2) |
| P | 3a | 0 | 0 | 0 | 1 | 0.0059(6) | 0.0059(6) | 0.0079(10) |
| Cu | 3b | 0 | 0 | 1/2 | 0.064(8) | 0.035(13) | 0.035(13) | 0.039(19) |

**FIG. S1**

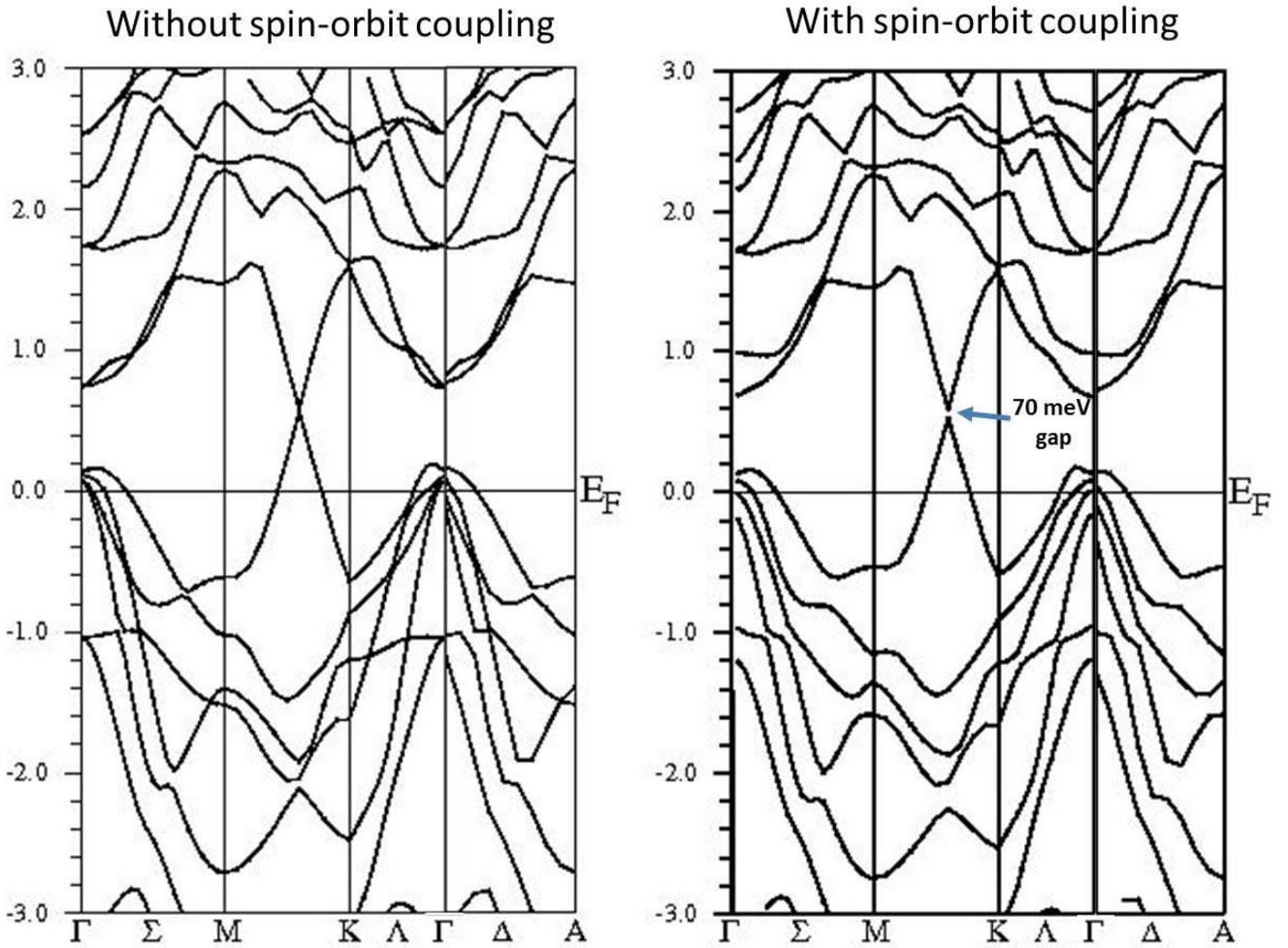

**FIG. S1** (Color online) Calculated band structures of $Zr_2Te_2P$ with (right) and without (left) spin-orbit coupling. A Dirac point is observed inside the pseudogap above $E_F$ without the consideration of spin-orbit coupling. However, a small gap of ~ 70 meV is opened when spin-orbit coupling is taken into account; even with the SOC the majority of these Dirac bands remain linear over a 2 eV energy range.

**FIG. S2**

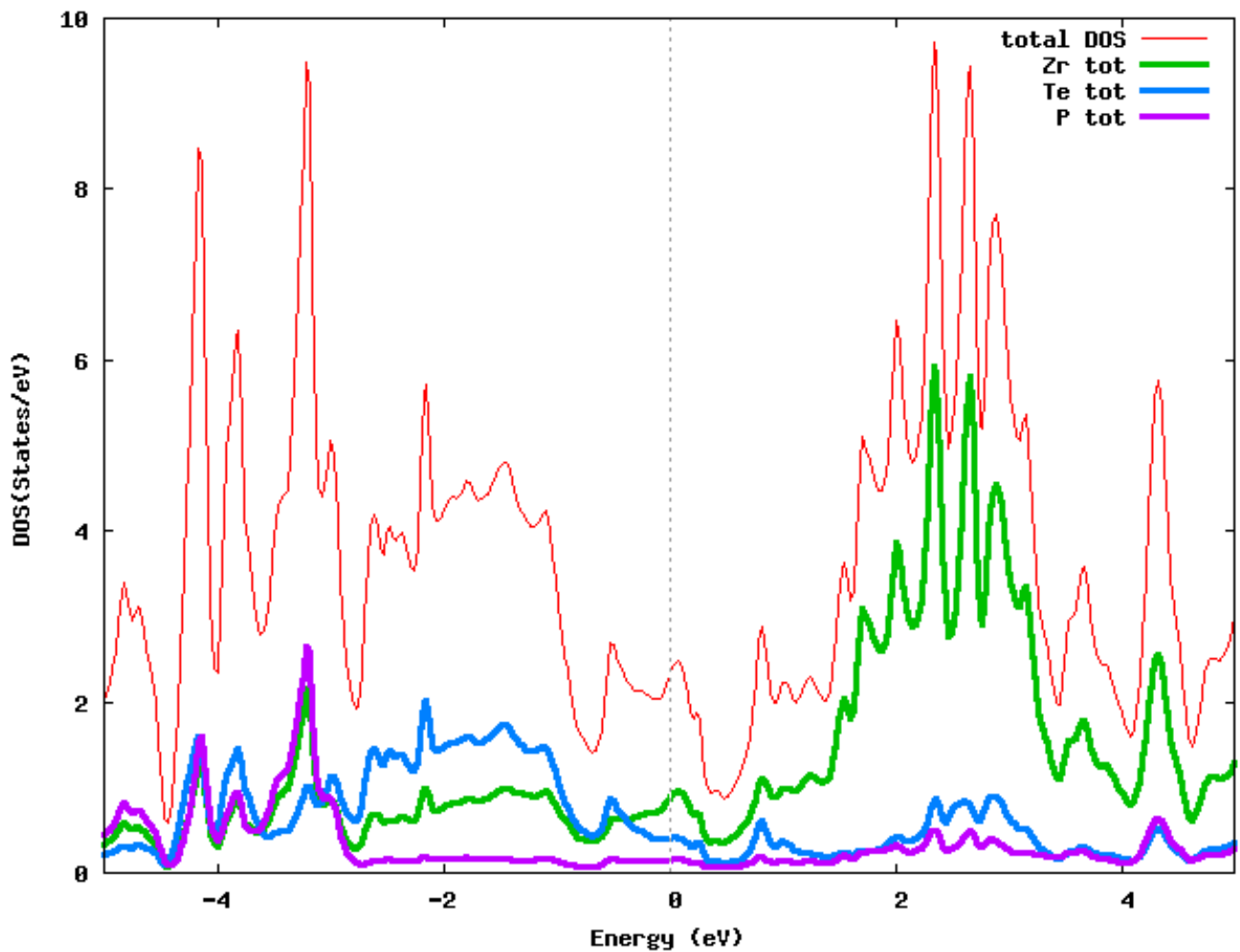

**FIG. S2** (Color online) Density of states calculation for $Zr_2Te_2P$. A pseudogap is observed at 0.2 – 0.7 eV above $E_F$.